\documentclass[usenatbib]{mn2e}
\usepackage{graphicx,natbib}
\usepackage{bm,url}

\newcommand{\EQ}{\begin{equation}}
\newcommand{\EN}{\end{equation}}
\newcommand{\EQA}{\begin{eqnarray}}
\newcommand{\ENA}{\end{eqnarray}}

\newcommand{\EEq}[1]{Equation~(\ref{#1})}
\newcommand{\Eq}[1]{Equation~(\ref{#1})}
\newcommand{\Eqs}[2]{Equations~(\ref{#1}) and~(\ref{#2})}

\newcommand{\App}[1]{Appendix~\ref{#1}}
\newcommand{\Sec}[1]{Section~\ref{#1}}
\newcommand{\Secs}[2]{Sections~\ref{#1} and \ref{#2}}
\newcommand{\Fig}[1]{Figure~\ref{#1}}

\newcommand{\Figs}[2]{Figures~\ref{#1} and \ref{#2}}
\newcommand{\Figss}[2]{Figures~\ref{#1}--\ref{#2}}
\newcommand{\Tab}[1]{Table~\ref{#1}}

\newcommand{\bra}[1]{\langle #1\rangle}

{}
{}
{}

{}
{}
{}
{}
{}
{}
{}
{}
{}
\newcommand{\meanBB}{\overline{\mbox{\boldmath $B$}}{}}{}
{}
{}
{}
{}
{}
{}
{}
{}
{}
{}
{}

{}
{}
{}

\newcommand{\meanB}{\overline{B}}

{}

{}
{}

\newcommand{\eee}{\hat{\mbox{\boldmath $e$}} {}}

\newcommand{\rr}{\mbox{\boldmath $r$} {}}
\newcommand{\yy}{\mbox{\boldmath $y$} {}}

\newcommand{\vv}{\mbox{\boldmath $v$} {}}

\newcommand{\kk}{\bm{k}}

\newcommand{\xx}{\bm{x}}

\newcommand{\BB}{\bm{B}}

\newcommand{\uu}{\mbox{\boldmath $u$} {}}

\newcommand{\JJ}{\mbox{\boldmath $J$} {}}

\newcommand{\AAA}{\mbox{\boldmath $A$} {}}

\newcommand{\ff}{\mbox{\boldmath $f$} {}}

\newcommand{\FF}{\mbox{\boldmath $F$} {}}

\newcommand{\nab}{\mbox{\boldmath $\nabla$} {}}

\newcommand{\oo}{\mbox{\boldmath $\omega$} {}}

\newcommand{\RRRR}{\mbox{\boldmath ${\sf R}$} {}}
\newcommand{\SSSS}{\mbox{\boldmath ${\sf S}$} {}}

\newcommand{\ii}{{\rm i}}

\newcommand{\DD}{{\rm D} {}}

\newcommand{\dd}{{\rm d} {}}

\def\la{\mathrel{\mathchoice {\vcenter{\offinterlineskip\halign{\hfil
$\displaystyle##$\hfil\cr<\cr\sim\cr}}}
{\vcenter{\offinterlineskip\halign{\hfil$\textstyle##$\hfil\cr<\cr\sim\cr}}}
{\vcenter{\offinterlineskip\halign{\hfil$\scriptstyle##$\hfil\cr<\cr\sim\cr}}}
{\vcenter{\offinterlineskip\halign{\hfil$\scriptscriptstyle##$\hfil\cr<\cr\sim\cr}}}}}
\def\ga{\mathrel{\mathchoice {\vcenter{\offinterlineskip\halign{\hfil
$\displaystyle##$\hfil\cr>\cr\sim\cr}}}
{\vcenter{\offinterlineskip\halign{\hfil$\textstyle##$\hfil\cr>\cr\sim\cr}}}
{\vcenter{\offinterlineskip\halign{\hfil$\scriptstyle##$\hfil\cr>\cr\sim\cr}}}
{\vcenter{\offinterlineskip\halign{\hfil$\scriptscriptstyle##$\hfil\cr>\cr\sim\cr}}}}}
\def\Pm{\mbox{\rm Pr}_{\rm M}}
\def\Rm{\mbox{\rm Re}_{\rm M}}

\def\Rmc{R_{\rm m}^{\rm crit}}
\def\Rey{\mbox{\rm Re}}

\def\EM{E_{\rm M}}
\def\EMz{E_{\rm M0}}
\def\HK{H_{\rm K}}
\def\HM{H_{\rm M}}

\def\cs{c_{\rm s}}

\def\kmean{k_{\rm m}}
\def\ksmall{k_{\rm s}}

\def\kf{k_{\rm f}}
\def\Bf{B_{\rm f}}
\def\epsf{\epsilon_{\rm f}}

\def\Brms{B_{\rm rms}}

\def\urms{u_{\rm rms}}

\def\etat{\eta_{\it t}}

\def\etaTz{\eta_{\rm T0}}

\def\etaT{\eta_{\rm T}}

\def\half{{\textstyle{1\over2}}}

\newcommand{\ynjp}[3]{ #1, {NJP,} {#2}, #3}
\newcommand{\yapj}[3]{ #1, {ApJ,} {#2}, #3}

\newcommand{\yapjl}[3]{ #1, {ApJ,} {#2}, #3}

\newcommand{\yan}[3]{ #1, {Astron.\ Nachr.,} {#2}, #3}

\newcommand{\yana}[3]{ #1, {A\&A,} {#2}, #3}

\newcommand{\yass}[3]{ #1, {Ap\&SS,} {#2}, #3}

\newcommand{\yjfm}[3]{ #1, {J.\ Fluid Mech.,} {#2}, #3}

\newcommand{\yjetp}[3]{ #1, {Sov.\ Phys.\ JETP,} {#2}, #3}

\newcommand{\yprl}[3]{ #1, {Phys.\ Rev.\ Lett.,} {#2}, #3}

\newcommand{\ymn}[3]{ #1, {MNRAS,} {#2}, #3}
\newcommand{\ynat}[3]{ #1, {Nature,} {#2}, #3}

\newcommand{\ypre}[3]{ #1, {Phys.\ Rev.\ E,} {#2}, #3}

\newcommand{\yjour}[4]{ #1, {#2}, {#3}, #4}

\newcommand{\ybook}[3]{ #1, {#2} (#3)}

\title[Large-scale dynamo action in the kinematic stage]{Traces of large-scale dynamo action in the kinematic stage}
\author[K. Subramanian \& A. Brandenburg]
{Kandaswamy Subramanian$^1$\thanks{E-mail:kandu@iucaa.ernet.in} and
Axel Brandenburg$^{2,3}$
\\
$^1$Inter University Centre for Astronomy and Astrophysics,
Post Bag 4, Pune University Campus, Ganeshkhind, Pune 411 007, India\\
$^2$Nordita, KTH Royal Institute of Technology and Stockholm University,
Roslagstullsbacken 23, SE-10691 Stockholm, Sweden\\
$^3$Department of Astronomy, AlbaNova University Center,
Stockholm University, SE-10691 Stockholm, Sweden
}

\date{\today,~ $ $Revision: 1.129 $ $}
\begin{document}

\maketitle

\begin{abstract}
Using direct numerical simulations (DNS) we verify
that in the kinematic regime, a turbulent helical dynamo
grows in such a way that the magnetic energy spectrum remains
to high precision shape-invariant, i.e., at each wavenumber $k$
the spectrum grows with the same growth rate.
Signatures of large-scale dynamo action can be identified through
the excess of magnetic energy at small $k$,
of one of the two oppositely polarized constituents.
Also a suitably defined planar average of the magnetic field can be
chosen such that its rms value isolates the strength of the mean field.
However, these different means of analysis suggest that the strength
of the large-scale field diminishes with increasing magnetic Reynolds
number $\Rm$ like $\Rm^{-1/2}$ for intermediate values and
like $\Rm^{-3/4}$ for larger ones.
Both an analysis from the Kazantsev model including helicity and the
DNS show that this arises due to the magnetic 
energy spectrum still peaking at resistive scales, even when helicity
is present.
As expected, the amplitude of the large-scale field increases with
increasing fractional helicity, enabling us to determine the onset
of large scale dynamo action and distinguishing it from that of the
small-scale dynamo.
Our DNS show that, contrary to earlier results for smaller scale separation
(only 1.5 instead of now 4), the small-scale dynamo can still be excited
at magnetic Prandtl numbers of 0.1 and only moderate values of the
magnetic Reynolds numbers ($\sim 160$).
\end{abstract}


\section{Introduction}
\label{intro}

The origin of large-scale magnetic fields in
astrophysical bodies such as stars and galaxies
remains an outstanding problem, given that
those fields are coherent on the scale of
the systems themselves.
Indeed, the observed scale is often larger
than the scale of the turbulent motions,
which would be the convective scale in the Sun
or the turbulent length scales induced
by supernova remnants in galaxies.
These large-scale magnetic fields are typically explained as being
due to turbulent dynamo action, whereby the combined action of helical
turbulence and shear amplifies and maintains fields coherent on scales
larger than the scales of random stirring.
We refer to this as the large-scale or mean-field dynamo.
However, when the magnetic Reynolds number, $\Rm$, is large,
such turbulent motions also generically lead to the
small-scale or fluctuation dynamo,
whereby magnetic fields coherent on scales of the order of or smaller
than the outer scales of the turbulence are rapidly generated.
In the following, we use mean-field and fluctuation dynamos
synonymously with large-scale and small-scale dynamos, respectively.

Typically, the growth rate of the fluctuation or small-scale dynamo
is much larger than the growth rate associated with the mean-field
or large-scale dynamo.
Then, in a system where both types of dynamos can in principle
operate, at least in the kinematic stage, 
magnetic fluctuations generated by the fluctuation dynamo
would in principle rapidly overwhelm the large-scale field
which could be generated by mean-field dynamo action.
The question then arises, whether in such a system there is any evidence
for large-scale fields at all in the kinematic stage. 

Large-scale dynamo action from helical turbulence has clearly been seen
in several direct numerical simulations (DNS)
during the late nonlinear stage when the magnetic
field is close to saturation \citep[e.g.][]{B01}.
This is partially due to the phenomenon of ``self-cleaning'',
which means the suppression of power on scales between
the largest and the driving scale of the turbulence.
However, during the early phase, there is no clear evidence for
large-scale dynamos, especially when small-scale dynamo action is
also expected to be possible.

Small-scale dynamo action is best studied in the case when there is no helicity
\citep[see][for a review]{BS05}.
In the presence of helicity, however, not only the large-scale
dynamo may become possible, but also the small-scale dynamo might get modified
such that large-scale and small-scale dynamos are just different aspects
of a single dynamo \citep{Sub99}.

It is instructive to think of the kinematic small-scale dynamo problem
as a quantum mechanical potential problem, where by the existence
of bound states in the potential, corresponds to growing modes of the
small-scale dynamo \citep{Kaz68}. An extension of this picture
in the presence of helicity is that the corresponding potential
allows for `tunnelling' of these bound states into `free-particle'
states \citep{Sub99,BS00,BCR05}. The larger growth rate
of the small-scale dynamo, compared to that of the large-scale dynamo,
is then reflected in the fact that the potential well at the scale, say $l$, 
where the bound state is located, 
is deeper than the scale where the free particle states exist, say $L$.
In case there is only a single fastest growing eigenfunction,
which grows fastest during the kinematic state, this change in the
potential depths at scales $l$ and $L$ could then reflect itself
in the corresponding strength of the eigenfunction, which would have
a larger amplitude on the scale $l$ than the scale $L$, or corresponding
wavenumbers proportional to $l^{-1}$ and $L^{-1}$.
Whether this picture is indeed a useful description
of the kinematic eigenfunction is currently unknown.

Our aim here is to examine whether in helical turbulence there is evidence for
the existence of the large-scale dynamo even in the presence of the
fluctuation dynamo.
To isolate features of the large-scale dynamo, we consider here,
for most part, the regime
of small magnetic Prandtl numbers, $\Pm=\nu/\eta$, where $\nu$ is the
kinematic viscosity and $\eta$ the magnetic diffusivity.
For small values of $\Pm$, e.g.\ for $\Pm=0.1$,
the small-scale dynamo is expected to be much
harder to excite if there were no helicity in the flow \citep{Isk07}.
The large-scale dynamo, on the other hand, is known to be virtually
independent of $\Pm$ and $\Rm$ once $\Rm>O(1)$; see \cite{B09} and \cite{MB10}.
One then expects this to provide a better chance of 
seeing evidence for the large-scale field in the kinematic stage.%
\footnote{This is reminiscent of ideas by \citet{TC13} and \citet{CT14}
where strong shear suppresses small-scale 
dynamo action and then allows large-scale dynamo waves to persist 
at high $\Rm$ in their helical flow models.}  
However as we will see below, even this small $\Pm$ case does not yield a 
decisive change, in preferentially hosting a large-scale dynamo. 

We restrict ourselves to the study of subsonic flows
with Mach numbers around 0.3.
While this is relevant to stars that also have small values of $\Pm$,
larger Mach numbers would be interesting and relevant to the study of
the warm and cold components of the interstellar medium, but this has
the problem that it results in the possibility of shocks.
This would force us to increase the viscosity, resulting in smaller
values of the Reynolds number.
It is well known that in supersonic flows, the small-scale dynamo
is harder to excite \citep{HBM04,Federrath11,Schober,Schleicher}, but the
large-scale dynamo, which is the subject of the present study,
depends essentially on the scale separation ratio of the turbulence
and may not (or only weakly) depend on the Mach number.
For example supernova-driven turbulence in galaxies, involving
flows at high Mach number,  
has been shown to be capable of driving a large-scale dynamo  
\citep{GZER08a,GZER08b,Gent13a,Gent13b}.

We begin by presenting the basic equations of our DNS (\Sec{Model}),
discuss then the results for different magnetic Reynolds and Prandtl
numbers (\Sec{Simulations}), and place them within the framework
of a unified analytical model (\Sec{unified}), before concluding in
\Sec{Conclusions}.

\section{Model}
\label{Model}

We consider dynamo action in a cubic domain of size $L_1^3$,
driven by turbulence forced at wavenumbers $\kf\approx4\,k_1$,
where $k_1=2\pi/L_1$ is the smallest wavenumber in the domain.
The forcing is assumed to be helical, so that one can in principle
have the operation of an $\alpha^2$ type large-scale dynamo.
To begin with, as explained above, we consider a small value of the
magnetic Prandtl number $\Pm=0.1$.

We solve the compressible hydromagnetic equations:
\begin{eqnarray}
&& \frac{\partial}{\partial t} \AAA = \uu\times\BB -\eta\mu_{0}\JJ,
\label{eq: induction} \\
&& \frac{\DD}{\DD t} \uu = -\cs^{2}\nab \ln{\rho} +
\frac{1}{\rho} \JJ\times\BB + \FF_{\rm visc} + \ff,
\label{eq: momentum} \\
&& \frac{\DD}{\DD t} \ln{\rho} = -\nab\cdot\uu, \label{eq: continuity}
\end{eqnarray}
where $\AAA$ is the magnetic vector potential, $\uu$ the velocity,
$\BB$ the magnetic field, $\eta$ the molecular magnetic diffusivity,
$\mu_{0}$ the vacuum permeability, $\JJ$ the electric
current density, $\cs$ the isothermal sound speed, $\rho$ the
density, $\FF_{\rm visc}$ the viscous force,
$\ff$ the helical forcing term,
and $\DD/\DD t = \partial/\partial t + \uu\cdot\nab$ the advective time
derivative.
The viscous force is given as
$\FF_{\rm visc} = \rho^{-1}\nab\cdot2\nu\rho\SSSS$,
where $\nu$ is the kinematic viscosity,
and $\SSSS$ is the traceless rate of strain tensor with components
${\sf S}_{ij}=\frac{1}{2}(u_{i,j}+u_{j,i})-\frac{1}{3}\delta_{ij}\nab\cdot\uu$.
Commas denote partial derivatives.

The energy supply for a helically driven dynamo is provided by the
forcing function $\ff = \ff(\xx,t)$, which is random in time
and defined as
\EQ
\ff(\xx,t)={\rm Re}\{N\ff_{\kk(t)}\exp[\ii\kk(t)\cdot\xx+\ii\phi(t)]\},
\label{ForcingFunction}
\EN
where $\xx$ is the position vector.
The wave vector $\kk(t)$ and the random phase
$-\pi<\phi(t)\le\pi$ change at every time step, so $\ff(\xx,t)$ is
$\delta$-correlated in time.
Therefore, the normalization factor $N$ has to be proportional to $\delta t^{-1/2}$,
where $\delta t$ is the length of the time step.
On dimensional grounds it is chosen to be
$N=f_0 c_{\rm s}(|\kk|c_{\rm s}/\delta t)^{1/2}$, where $f_0$ is a
nondimensional forcing amplitude.
We choose $f_0=0.02$, which results in a maximum Mach number of about 0.3
and an rms velocity of about 0.085, which is almost the same for all the runs.
At each timestep we select randomly one of many possible wave vectors
in a certain range around a given forcing wave number with
average value $k_{\rm f}$.
Transverse helical waves are produced via \citep{HBD04}
\begin{equation}
\ff_{\kk}=\RRRR\cdot\ff_{\kk}^{\rm(nohel)}\quad\mbox{with}\quad
{\sf R}_{ij}={\delta_{ij}-\ii\sigma\epsilon_{ijk}\hat{k}_k
\over\sqrt{1+\sigma^2}},
\label{eq: forcing}
\end{equation}
where $\sigma$ is a measure of the helicity of the forcing and
$\sigma=1$ for positive maximum helicity of the forcing function
and
\EQ
\ff_{\kk}^{\rm(nohel)}=
\left(\kk\times\eee\right)/\sqrt{\kk^2-(\kk\cdot\eee)^2}
\label{nohel_forcing}
\EN
is a nonhelical forcing function, where $\eee$ is an arbitrary unit vector
not aligned with $\kk$; note that $|\ff_{\kk}|^2=1$ and
\EQ
\ff_{\kk}\cdot(\ii\kk\times\ff_{\kk})^*=2\sigma k/(1+\sigma^2),
\EN
so the relative helicity of the forcing function in real space is
$2\sigma/(1+\sigma^2)$.

Our model is governed by several nondimensional parameters.
In addition to the scale separation ratio $\kf/k_1$, introduced above,
there are the magnetic Reynolds and Prandtl numbers
\EQ
\Rm=\urms/\eta\kf,\quad
\Pm=\nu/\eta.
\label{Rey_def}
\EN
These two numbers also define the fluid Reynolds number,
$\Rey=\urms/(\nu\kf)=\Rm/\Pm$.
The maximum values that can be attained are limited by the numerical
resolution and become more restrictive at larger scale separation.
The calculations have been performed using the {\sc Pencil Code}\footnote{
\url{http://pencil-code.googlecode.com}} at resolutions between $128^3$
and $1024^3$ mesh points.

\section{Simulations}
\label{Simulations}

In the following we present runs at different values
of $\Rm$, $\Pm$, and $\sigma$; see \Tab{Tsummary}.

\begin{table}\caption{
Summary of runs discussed in this paper.
}\vspace{12pt}\centerline{\begin{tabular}{lrlcrr}
\hline
\hline
Run & $\;\Rm\!\!$ & $\!\!\Pm\;$ & $\sigma$ & $\!\!\!\tilde\lambda\quad$
    & $N\;\;$ \\
\hline
A01 &  1  &  0.1  &  1  &  $-0.004$ & $128^3$ \\ 
B01 &  3  &  0.1  &  1  &  $ 0.014$ & $128^3$ \\ 
C01 & 16  &  0.1  &  1  &  $ 0.029$ & $128^3$ \\ 
D01 & 33  &  0.1  &  1  &  $ 0.033$ & $256^3$ \\ 
E01 & 65  &  0.1  &  1  &  $ 0.036$ & $256^3$ \\ 
F01 &160  &  0.1  &  1  &  $ 0.038$ & $256^3$ \\ 
F02 &160  &  0.2  &  1  &  $ 0.038$ & $256^3$ \\ 
F05 &160  &  0.5  &  1  &  $ 0.041$ & $256^3$ \\ 
F07 &160  &  0.7  &  1  &  $ 0.045$ & $256^3$ \\ 
F1  &160  &   1   &  1  &  $ 0.051$ & $256^3$ \\ 
G01 &340  &  0.1  &  1  &  $ 0.040$ &$1024^3$ \\ 
G02 &360  &  0.2  &  1  &  $ 0.037$ &$1024^3$ \\ 
G05 &330  &  0.5  &  1  &  $ 0.050$ & $512^3$ \\ 
G1  &330  &   1   &  1  &  $ 0.069$ & $256^3$ \\ 
\hline
\hline
F01b&160  &  0.1  & 0.7 &  $ 0.032$ & $256^3$ \\ 
F01c&160  &  0.1  & 0.5 &  $ 0.023$ & $256^3$ \\ 
F01d&160  &  0.1  & 0.3 &  $ 0.010$ & $256^3$ \\ 
F01e&160  &  0.1  & 0.2 &  $ 0.005$ & $512^3$ \\ 
F01f&160  &  0.1  & 0.1 &  $ 0.003$ & $512^3$ \\ 
\hline
\hline
f005&160  &  0.05 &  0  &  $ 0.001$ & $512^3$ \\ 
g005&310  &  0.05 &  0  &  $ 0.015$ & $512^3$ \\ 
\hline
f01 &160  &  0.1  &  0  &  $ 0.003$ & $512^3$ \\ 
g01 &200  &  0.1  &  0  &  $ 0.006$ & $512^3$ \\ 
\hline
e02 & 80  &  0.2  &  0  &  $-0.003$ & $256^3$ \\ 
f02 &160  &  0.2  &  0  &  $ 0.015$ & $256^3$ \\ 
\hline
d05 & 30  &  0.5  &  0  &  $-0.004$ & $128^3$ \\ 
e05 & 80  &  0.5  &  0  &  $ 0.016$ & $128^3$ \\ 
\hline
d1  & 40  &   1   &  0  &  $ 0.010$ & $128^3$ \\ 
e1  & 60  &   1   &  0  &  $ 0.019$ & $128^3$ \\ 
f1  &150  &   1   &  0  &  $ 0.045$ & $128^3$ \\ 
\label{Tsummary}\end{tabular}}\end{table}

\subsection{Growth rate}

\begin{figure}\begin{center}
\includegraphics[width=\columnwidth]{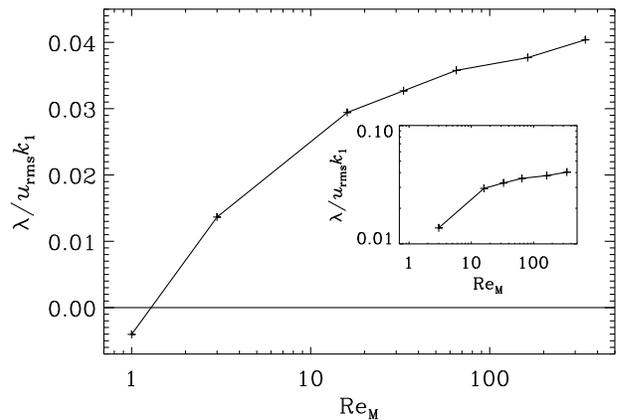}
\end{center}\caption[]{
Normalized growth rate versus $\Rm$ for $\Pm=0.1$ and $\sigma=1$.
For intermediate values of $\Rm$ below 100, the growth rate corresponds
to that of a helical large-scale dynamo.
The inset shows $\lambda/\urms\kf$ vs.\ $\Rm$ in double-logarithmic
representation.
}\label{pplam_vs_Rm}\end{figure}

It turns out that for helical driving, and $\Pm=0.1$,
the onset of dynamo action occurs at
small values of $\Rm$; see \Fig{pplam_vs_Rm},
where we show the normalized growth rate, $\lambda/\urms\kf$,
of a dynamo as a function of $\Rm$.
We see that, for $\kf/k_1=4$, the critical value of $\Rm$ is around 2.
Furthermore, the increase of $\lambda$ becomes less steep for
$\lambda/\urms\kf\ga0.03$,
which is a value that was found earlier
for fully helical large-scale dynamos \citep{B09},
who also used $\kf/k_1=4$.

\begin{figure}\begin{center}
\includegraphics[width=\columnwidth]{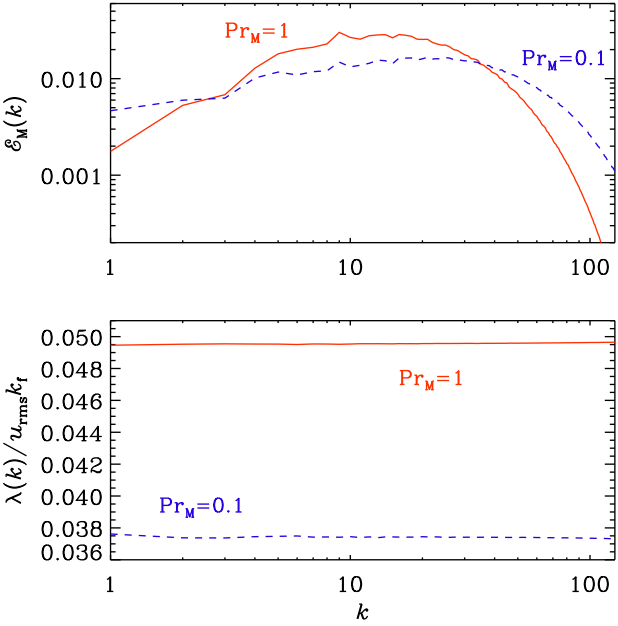}
\end{center}\caption[]{
Spectrum of magnetic energy during the kinematic phase for 
$\Pm=0.1$ (Run~F01; blue, dashed lines, $B_{\rm ini}\approx4\times10^{-31}$)
and $\Pm=1$ (Run~F1; red, solid lines, $B_{\rm ini}\approx2\times10^{-35}$)
using $\sigma=1$ in both cases.
The corresponding growth rates as a function of $k$ are given
in the bottom panel.
}\label{ppspecmtfit_comp0}\end{figure}

\subsection{Wavenumber-dependent growth rate}

One of the features that we want to examine is whether the magnetic
field grows as an eigenfunction in the kinematic stage, when both
large- and small-scale  dynamo action is possible.
For this we look at the time evolution of magnetic energy spectra,
$\EM(k,t)$.
It is convenient to represent the time evolution in the form
\begin{equation}
\EM(k,t)=\EMz(k)\,e^{2\lambda(k)t}.
\label{fit}
\end{equation}
Since $\EMz(k)$ depends on the initial magnetic field
strength, $B_{\rm ini}$, it is convenient to write it as
\begin{equation}
\EMz(k)=\half B_{\rm ini}^2{\cal E}_{\rm M}(k),
\end{equation}
where ${\cal E}_{\rm M}(k)$ is the normalized spectrum with
$\int {\cal E}_{\rm M}(k)\,\dd k=1$.
Note that we have here allowed for a $k$-dependent growth rate, $\lambda(k)$.
This enables us to assess quantitatively to what extent the growth rate
depends on $k$.
The resulting $\lambda(k)$ is shown in the bottom panel of
\Fig{ppspecmtfit_comp0} for $\Pm=0.1$ and $\Pm=1$, respectively.
We see that, to very good accuracy, the growth rate is the same
for different wavenumbers, confirming that the spectra grow as 
one eigenfunction, even when both large-scale and small-scale dynamos
are possible, due to helical forcing. 

\begin{figure}\begin{center}
\includegraphics[width=\columnwidth]{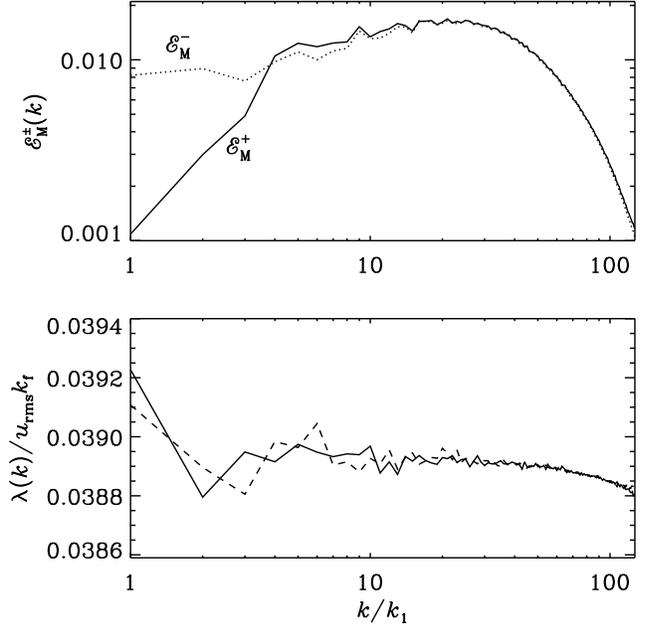}
\end{center}\caption[]{
Spectrum of positively and negatively polarized contributions during
the kinematic phase for Run~F01 with $\sigma=1$, $\Pm=0.1$, and $\Rm\approx160$.
The growth rate is given separately for the spectra of magnetic energy
of positively (solid line) and negatively (dotted line) polarized contributions.
}\label{ppspecmtfit_comp}\end{figure}

\begin{figure}\begin{center}
\includegraphics[width=\columnwidth]{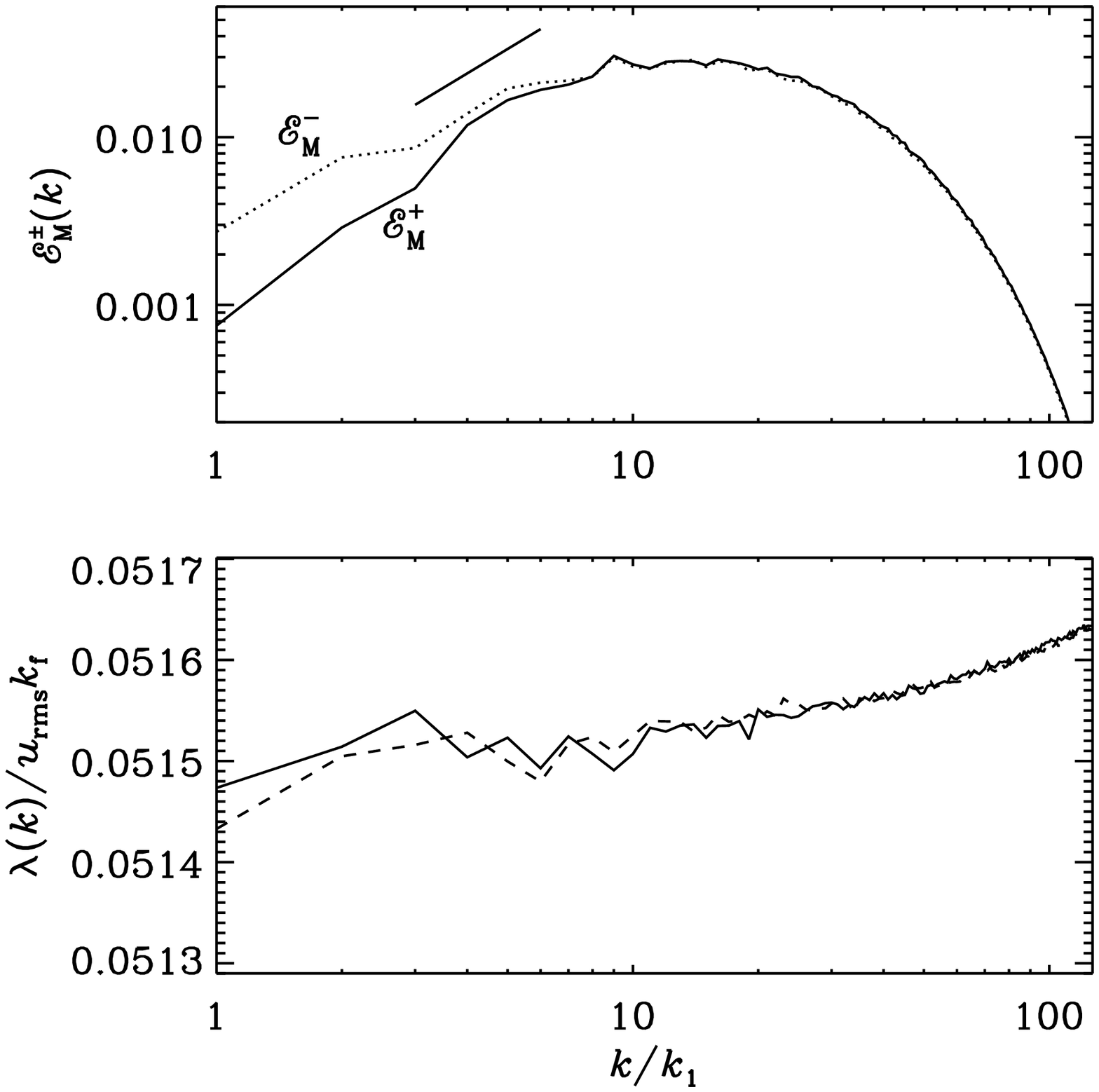}
\end{center}\caption[]{
Same as \Fig{ppspecmtfit_comp},
but for Run~F1 with $\sigma=1$, $\Pm=1$,
$\Rm\approx160$, and $\Pm=1$.
The short straight line gives the $k^{3/2}$ Kazantsev slope for orientation.
}\label{ppspecmtfit_comp_256Pm1a}\end{figure}

\begin{figure}\begin{center}
\includegraphics[width=\columnwidth]{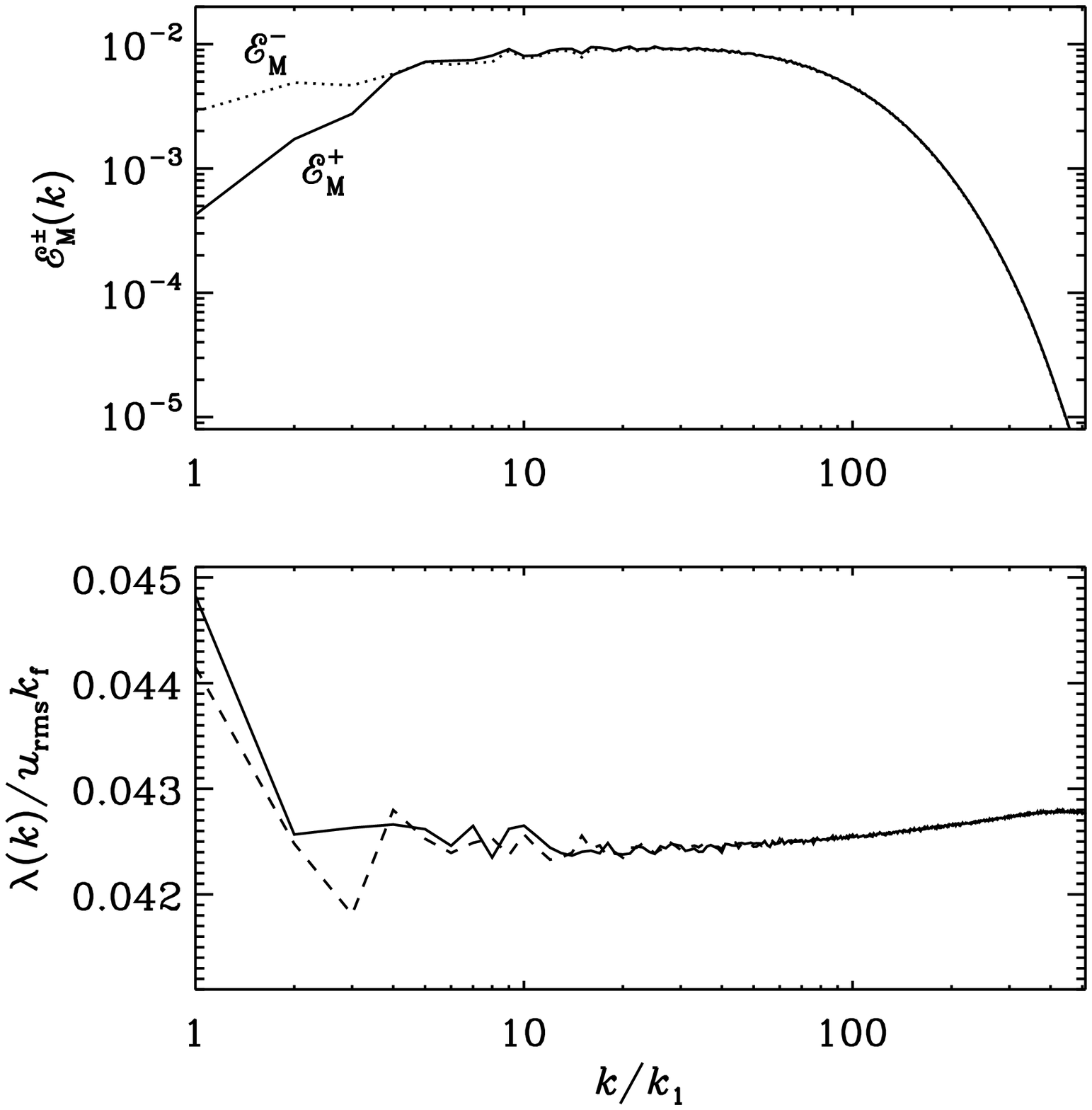}
\end{center}\caption[]{
Same as \Fig{ppspecmtfit_comp},
but for Run~G01 with $\sigma=1$, $\Pm=0.1$, and $\Rm\approx330$.
}\label{ppspecmtfit_comp_L1024Pm01a}\end{figure}

\subsection{Magnetic spectra in the polarization basis}
\label{MagneticSpectra}

For the $\alpha^2$ dynamo, which arises in helical turbulence,
due to magnetic helicity conservation, one expects the helicity
of small-scale and large-scale fields to have different signs
at early times.
Thus, one would be able to see a clearer
signature of the large-scale field, if one looks separately for
positively and negatively polarized helical fields, defined as 
\begin{equation}
\EM^\pm(k,t)=\half\left[\EM(k,t)\pm\half k\HM(k,t)\right].
\end{equation}
Again, we fit the resulting spectra to an exponential growth,
analogous to \Eq{fit}, and plot the normalized magnetic energy spectra
${\cal E}_{\rm M}^\pm(k)$.
They are are shown in the top panels of 
\Fig{ppspecmtfit_comp} for $\Pm=0.1$,
and \Fig{ppspecmtfit_comp_256Pm1a} for $\Pm=1$.
We see that there is indeed excess power in $\EMz^-$ at small $k$
corresponding to the large-scale field generated in such
helical turbulence.
For Run~F1 with $\Pm=1$, there is also a short range with
Kazantsev $k^{3/2}$ scaling.
On the other hand, for $\Pm=0.1$ the scaling is significantly
flatter, as can be seen from \Fig{ppspecmtfit_comp_L1024Pm01a},
where we show the result for Run~G01 with $\Rm\approx330$.
Since $\Pm=0.1$, we have here $\Rey=3300$.
Note that there is a small uprise of $\lambda(k)$ at $k=k_1$,
which may however be a consequence of the time interval $\Delta t$
being too short (here, $\lambda\Delta t=10$,
while in all other cases it is at least 30).

We can see from \Figss{ppspecmtfit_comp}{ppspecmtfit_comp_L1024Pm01a}
that the magnetic energy spectra rise with $k$, and peak at wavenumbers 
much larger than the forcing wavenumber $\kf$, and closer
to the resistive scale. Therefore, even though
there is clear evidence for excess power corresponding
to the large-scale field, the rms field is likely to be
dominated by small scales, perhaps close to the resistive scale.
We will return to this aspect of the kinematic dynamo below.

\subsection{Expectation from $\alpha^2$ dynamos}

It is useful to compare the wavenumber of where excess power
would occur in an $\alpha^2$ dynamo.
In such a model, the mean magnetic field $\meanBB$ is governed
by the equation
\begin{equation}
{\partial\meanBB\over\partial t}=\nab\times(\alpha\meanBB)
+\etaT\nabla^2\meanBB,
\end{equation}
where $\alpha$ characterizes the strength of the $\alpha$ effect
and $\etaT=\eta+\etat$ is the sum of microphysical and turbulent
magnetic diffusivities.
Solutions proportional to $\exp(\ii\kk\cdot\xx+\lambda t)$
give the growth rate as $\lambda=|\alpha k|-\etaT k^2$.
Its maximum value is attained when $\dd\lambda/\dd k=0$,
giving the peak at $k_{\rm peak}=|\alpha|/2\etaT$.
Based on results of the second order correlation approximation
applied to the high-conductivity limit \citep{KR80}, one has
$\alpha\approx\tau\bra{\oo\cdot\uu}/3$ and
$\etat\approx\tau\bra{\uu^2}/3$, where $\oo=\nab\times\uu$
is the vorticity of the small-scale turbulent flow $\uu$
and $\tau$ is the correlation time.
For maximally helical flows we expect $|\alpha|\approx\urms/3$ and
$\etat\approx\urms/3\kf$, so $k_{\rm peak}\approx\kf/2$ \citep{BDS02}.
Thus, the theoretically expected scale separation is only a factor of two.
This explains that it is in general difficult to identify excess power
at the wavenumber $\kf/2$ of the $\alpha^2$ dynamo compared with the
wavenumber $\kf$ of the turbulence.

Furthermore, the growth rate of the $\alpha^2$ dynamo
is given by substituting $k_{\rm peak}$ into the
above expression for $\lambda$. We get $\lambda=\lambda_{\rm peak}
= |\alpha|^2/4\etaT \sim \urms\kf/12\approx0.08\urms\kf$.
This can be compared
with the growth rate obtained in the DNS of $\lambda \sim 0.038\urms\kf$ for
$\Pm=0.1$, $\Rm=160$ case to $\lambda \sim 0.051\urms\kf$ for $\Pm=1$ case.
The smaller value obtained in the DNS perhaps
indicates that the field grows less efficiently.

\subsection{Growth of planar averages}

Another way to isolate the large-scale mean field is to
consider horizontal averages of the total magnetic field.
We define mean fields as one of three possible planar averages,
and determine their rms fields, denoted by
\begin{eqnarray}
\meanB^X=\bra{\bra{\BB}_{yz}^{2}}_x^{1/2},\cr
\meanB^Y=\bra{\bra{\BB}_{xz}^{2}}_y^{1/2},\cr
\meanB^Z=\bra{\bra{\BB}_{xy}^{2}}_z^{1/2}.
\end{eqnarray}
Here, the subscripts behind angle brackets denote the direction
over which the average is taken and the capital letter superscript
on $\meanB$ indicates the direction in which the mean field varies.
These averages allow one to isolate the rms values of the
eigenfunctions of the $\alpha^2$ dynamo.
The average relevant for our considerations is the one
that produces the largest rms value.
Which of the three averages it is, is a matter of chance,
because the system is statistically isotropic.

\begin{figure}\begin{center}
\includegraphics[width=\columnwidth]{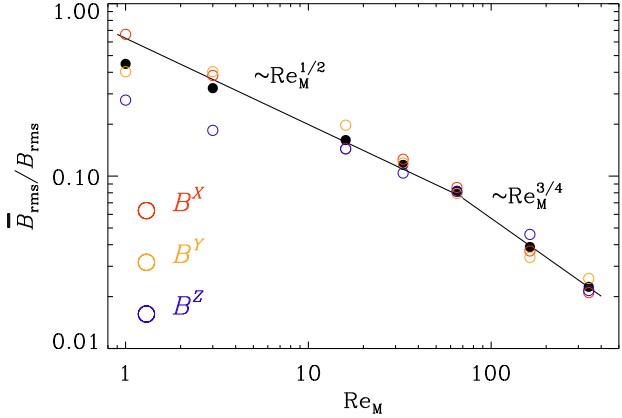}
\end{center}\caption[]{
Root-mean-squared value of the mean field relative to that of
the total field versus $\Rm$ for $\Pm=0.1$ and $\sigma=1$
during the kinematic stage.
The filled circles denote the averaged DNS results as an
average over the contributions from $\meanB^X$, $\meanB^Y$, and $\meanB^Z$.
The straight lines correspond to $0.63\,\Rm^{-1/2}$ and
$1.8\,\Rm^{-3/4}$ for lower and larger values of $\Rm$, respectively.
}\label{pbm}\end{figure}

In \Fig{pbm}, we show the ratios of the strength of the three mean fields,
$\meanB^X$, $\meanB^Y$, and $\meanB^Z$ defined above, to the total rms field
as a function of $\Rm$, for the case $\Pm=0.1$ and $\sigma=1$.
We see a fairly strong mean field for $\Rm\approx1$, but
as we increase $\Rm$, the fractional contribution of the large-scale
field during the kinematic phase decreases
proportional to $\Rm^{-1/2}$; see \Fig{pbm}.
For large values of $\Rm$, the scaling becomes even steeper.
In other words, the magnetic energy of the mean field decreases
inversely proportional to $\Rm$.
Similar scalings for the energy of the mean magnetic field were sometimes
expected to occur in the nonlinear stage \citep{VC92}, but here
it is a property of the dynamo in the linear regime.

We recall that for the present case of a homogeneous $\alpha^2$ dynamo
with periodic boundary conditions the saturation energy is indeed
independent of $\Rm$, although the time scale on which such as state
is reached increases with time proportional to $\Rm$ \citep{B01,CB13}.
We will return to the question of the decreasing strength of the
mean magnetic field during the linear stage in \Sec{unified},
where we will examine the solutions of the
Kazantsev model, generalized to include a helical velocity field
\citep{Sub99,RK99,BS00,BCR05,MB07,MB10}. 

\subsection{Dependence on fluid Reynolds number}

\begin{figure}\begin{center}
\includegraphics[width=\columnwidth]{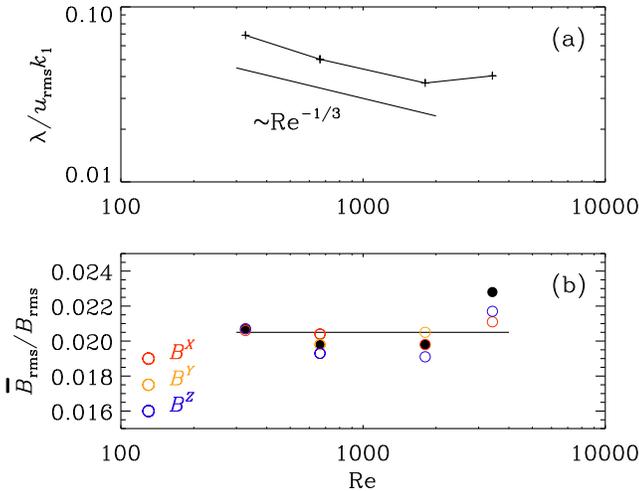}
\end{center}\caption[]{
(a) Normalized growth rate versus $\Rey$ for $\Rm\approx330$ and $\sigma=1$
and (b) root-mean-squared value of the mean field relative to that of
the total field during the kinematic stage.
Similar to \Fig{pbm}, the filled circles denote the averaged DNS results as an
average over the contributions from $\meanB^X$, $\meanB^Y$, and $\meanB^Z$.
The horizontal line is shown for reference.
}\label{pbm_Rm330}\end{figure}

It is well known that for large $\Pm$ ($ \gg 1$),
the growth rate of the small-scale dynamo 
scales with $\Rey$ \citep[asymptotically like $\Rey^{1/2}$; see][]{Scheko04}
and is independent of $\Rm$.
This is because for $\Pm\gg1$, the growth rate scales
with the eddy turn-over rate at the viscous scale, which increases
with $\Rey$. On the other hand, in the case of small $\Pm \ll 1$,
the growth rate scales as the eddy turnover rate at the
resistive scale, and hence as $\Rm^{1/2}$ \citep{MB10}.
We may now ask what happens for fully
helical flows with $\sigma=1$. 
This is shown in
\Fig{pbm_Rm330}(a), where we show the dependence of $\lambda$
on $\Rey$ for $\Rm\approx330$ (Runs~G01--G1).
Instead, we see actually a weak decline with increasing $\Rey$.
Furthermore, the fractional strength of the mean field
stays fixed; see \Fig{pbm_Rm330}(b).

\subsection{Fractional helicity}

As shown above, the onset of large-scale dynamo action occurs
at rather small values of $\Rm$, but it does require the presence
of helicity in the flow.
Therefore, the onset of large-scale dynamo action is mainly determined by
the amount of helicity, which is quantified by the dynamo number.
For an $\alpha^2$ dynamo, the relevant dynamo number is
$C_\alpha=\alpha/\etaTz k_1$, but in DNS this quantity is well
approximated by the quantity \citep{BB02,CB13}
\begin{equation}
C_\alpha^{\rm DNS}=\epsf\kf/k_1,
\end{equation}
where $\kf/k_1$ is the scale separation ratio,
$\epsf=\bra{\oo\cdot\uu}/\kf\urms^2$ is the fractional helicity,
$\oo=\nab\times\uu$ is the vorticity, and $\urms$ the rms velocity
of the turbulence.

In \Fig{pplam_vs_sig} we show $\lambda$ versus $\sigma$ and $\epsf$.
Note that $\epsf\approx2\sigma/(1+\sigma^2)$ is obeyed to a good
approximation \citep{CB13}.
We see that there is an imperfect bifurcation at $\epsf\approx0.3$.
For large-scale dynamo action to be possible, 
one needs $C_\alpha > 1$ which requires $\epsf > k_1/\kf =0.25$. 
The value $\epsf\approx0.3$
obtained here is slightly above this theoretical minimum.
If one wanted to capture the large-scale dynamo for even smaller
$\epsf$, then one requires a smaller $k_1/\kf$, which implies
either a larger box size or a smaller forcing scale.

\subsection{Transition to small-scale dynamos}

\begin{figure}\begin{center}
\includegraphics[width=\columnwidth]{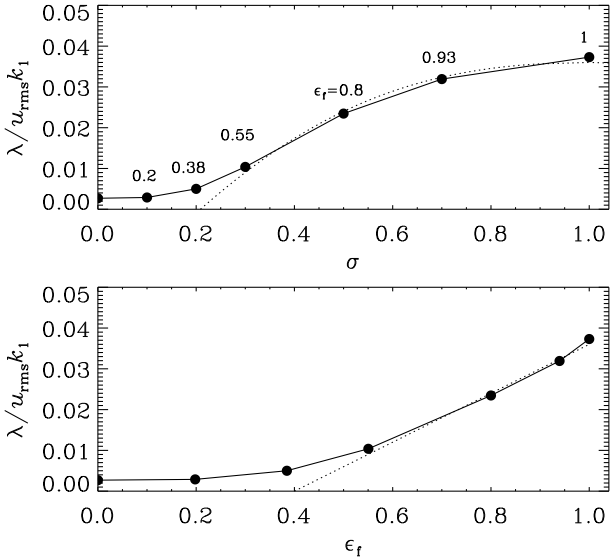}
\end{center}\caption[]{
Normalized growth rate versus $\sigma$ (upper panel,
$\epsf$ indicated on the data points)
and $\epsf$ (lower panel), for $\Pm=0.1$ and $\Rm\approx160$.
The dotted line indicates the tangent and thereby the
approximate position of the bifurcation line
if the bifurcation was a perfect one.
}\label{pplam_vs_sig}\end{figure}

Contrary to earlier findings for non-helical turbulence
driven at the scale of the domain ($\kf\approx1.5$ as opposed to the
value 4 used here), the small-scale dynamo is excited even for $\Pm=0.1$.
This can be seen from the fact that $\lambda > 0$ even when $\epsf=0$;
see \Fig{pplam_vs_sig}.
\cite{Scheko05} were unable to find small-scale dynamo solutions for
$\Pm=0.1$ and later \cite{Isk07} found negative growth rates at $\Pm=0.1$,
but positive values for $\Pm=0.05$.
This nonmonotonic behaviour was associated with the existence of a bottleneck
in the kinetic energy spectrum, i.e., a shallower spectrum near the
viscous sub-range, where the small-scale dynamo operates.
In the nonlinear regime, however, no such nonmonotonic behaviour is seen
\citep{B11}.

\begin{figure}\begin{center}
\includegraphics[width=\columnwidth]{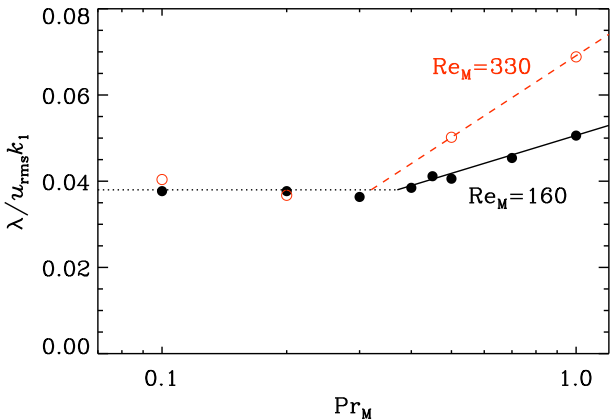}
\end{center}\caption[]{
Normalized growth rate versus $\Pm$ for $\sigma=1$,
$\Rm\approx160$ (Runs~F01--F1; solid black line with filled symbols) or
$\approx330$ (Runs~G01--G1; dashed red line with open symbols)
and varying values of $\Rey=\Rm/\Pm$.
The dotted line indicates the growth rate for a predominantly
large-scale dynamo.
}\label{pplam_vs_Pm}\end{figure}

As we increase $\Rm$, the small-scale dynamo becomes more strongly
supercritical and the critical value of $\Pm$ decreases from 0.4 to 0.3
as we increase $\Rm$ from 160 to 330; see \Fig{pplam_vs_Pm}.
Of course, for the fully helical case of this figure,
even for $\Pm=0.1$, 
the dynamo is really a combination of both the large-scale and 
small-scale dynamos, as we discussed in relation to \Fig{pplam_vs_sig}.
In addition, \Fig{pplam_vs_Pm} suggests 
that the behaviour of the dynamo changes from
a mainly large-scale dynamo at small $\Pm$ to one that becomes
even more
strongly controlled by small-scale dynamo action at larger $\Pm$.

\subsection{$\Rmc$ for the small-scale dynamo at low $\Pm$}

Early DNS of small-scale dynamos have focussed on homogeneous
turbulence in a periodic domain where random forcing was applied at the
scale of domain, so the forcing wavenumber was typically between 1 and 2
\citep{HBD04,Scheko04}.
In that case the critical value of $\Rm$ increased beyond 400
\citep{Scheko05}, but decreased again for smaller values of $\Pm$
\citep{Isk07}, which was argued to be a consequence of the bottleneck
effect in the kinetic energy spectrum near wavenumber where the
small-scale dynamo grows fastest.
In nonlinear simulations, on the other hand, the bottleneck effect
is suppressed and nonlinear small-scale dynamo action is sustained
at $\Pm=0.1$ for $\Rm\ga160$.

\begin{figure}\begin{center}
\includegraphics[width=\columnwidth]{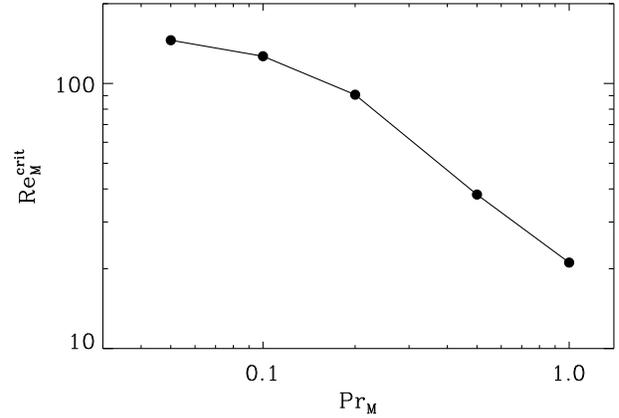}
\end{center}\caption[]{
Critical magnetic Reynolds number as a function of magnetic Prandtl number
for $\kf=4$ and $\sigma=0$ obtained by interpolating the growth rates
of Runs~f005--f1 in \Tab{Tsummary}.
}\label{pRm_crit}\end{figure}

Our new work now suggests that this might have been an artefact
of an artificially small forcing wavenumber.
Our new DNS with a forcing wavenumber $\kf=4\,k_1$ suggest
that small-scale dynamo action is excited at the usual values of $\Rm$
even when $\Pm=0.1$; see \Fig{pRm_crit}.
The reason for this lies probably in the fact that the bottleneck effect
is now weaker and that it is connected with particular issues related
to the way turbulence is driven.
We note that the increase of $\Rmc$ with decreasing $\Pm$ (\Fig{pRm_crit}) is
qualitatively similar to that obtained from the Kazantsev model
by \citet{MB10}; see the $h=0$ curve in their Fig~2.

\section{Interpretation in terms of the Kazantsev model with helicity}
\label{unified}

In order to interpret and further enhance the results from
the DNS, it is instructive to look at the
Kazantsev model with helicity \citep{VK86,Sub99,BS00,BS05,BCR05,MB07,MB10}.
In this model, the velocity is assumed to be a
statistically isotropic, homogeneous random field, and
$\delta$-correlated in time.
The two-point spatial correlation function of the
velocity field can be written as
$\bra{v_i({\xx},t)v_j({\yy},s)} = T_{ij}(r)\delta (t-s)$, where 
$r=|\rr|$ with $\rr = \xx - \yy$ and 
\EQ
T_{ij}(r) =
\left(\delta_{ij}-{r_i r_j \over r^2}\right)\,T_{N}
+{r_i r_j \over r^2}\,T_{L}
+\epsilon_{ijk} r_k\,F.
\label{tijhel}
\EN
Here $\bra{\cdot}$ denotes averaging over an ensemble of the stochastic velocity
field $\vv$, and we have written the correlation function 
in a form appropriate for a statistically
isotropic and homogeneous tensor \citep[cf.\ Section~34 of][]{landau}.
In \Eq{tijhel}, $T_{\rm L}(r)$, $T_{\rm N}(r)$ and $F(r)$ are the
longitudinal, transverse and helical parts of the 
correlation function for the velocity field. 
For an incompressible velocity field
$T_{\rm N} = (1/2r) [d(r^2 T_{\rm L})/dr]$.
The magnetic field $\BB$ is also assumed to be a 
statistically isotropic, homogeneous random field. 
Its equal-time, 
two-point correlation, $M_{ij}(r,t)$, is given by
\begin{equation}
M_{ij} = \left(\delta_{ij}-{r_i r_j \over r^2}\right)\,M_{\rm N}
+{r_i r_j\over r^2}\,M_{\rm L}
+\epsilon_{ijk} r_k\,C,
\label{mcor}
\end{equation}
where $M_L(r,t)$ and $M_{N}(r,t)$ are the longitudinal and transverse
correlation functions of the field, and 
$C(r,t)$ represents
the contribution from current helicity to the two-point correlation.
Since $\nabla\cdot\BB =0$, 
$M_{N}(r,t) = (1/2r) [\partial (r^2 M_L)/\partial r]$.
Using the induction equation, 
the evolution equations for $M_L(r,t)$ and $C(r,t)$ are given by
\citep{VK86,Sub99,BS00,BS05}, 
\EQ
{\partial M_{\rm L}\over\partial t} = {2\over r^4}{\partial \over \partial r}
\left[r^4 \eta_{\rm T} {\partial M_{\rm L} \over \partial r}\right]
+ 2G M_{\rm L} + 4\alpha C,
\label{mlequn}
\EN
\EQ
{\partial H \over \partial t}=-2\eta_{\rm T}C+ \alpha M_{\rm L}, \quad 
C = - \left(H'' + {4H' \over r}\right),
\label{mhequn}
\EN
where primes denote $r$ derivatives and
$\eta_{\rm T}(r)=\eta+\eta_{\rm t}(r)$ is the sum of
the microscopic diffusivity $\eta$ and an effective
scale-dependent turbulent magnetic diffusivity
$\eta_{\rm t}(r) = T_{\rm L}(0) - T_{\rm L}(r)$.
The term $G = -2(T_{\rm L}^{\prime\prime} + 4 T_{\rm L}^{\prime}/r)$
characterizes the rapid generation of magnetic fields by velocity shear
and $\alpha(r) = -2[F(0) - F(r)]$
represents the effect of kinetic helicity on the magnetic field.
It is related to the usual $\alpha$ effect in mean-field electrodynamics
\citep{Mof78}, but it is scale-dependent as in \cite{Mof83} and \cite{BRS08}.

\subsection{Bound states and tunnelling}

It is worth recalling some well-known properties of this system;
cf.\ \cite{BS05} and references therein.
In the absence of $F(r)$, the system describes the
fluctuation or small-scale dynamo.
Assuming solutions to be proportional to $\exp(\lambda t)$,
the evolution equation for $M_L$
can be transformed to a Schr\"odinger-type equation, with 
a potential $U(r)$ depending on $T_L$, and an energy eigenvalue
$E = -\lambda$.
Thus, {\it bound states} in the potential $U$
correspond to growing solutions with $\lambda > 0$.
This potential is positive with $U \to 2\eta/r^2 > 0$ as $r \to 0$,
while $U \to 2\etaTz/r^2 > 0$ as $ r\to \infty$,
when $T_{\rm L}(r) \to 0$. 
Here $\etaTz= \eta + T_{L}(0)$ is the sum of microscopic
and turbulent diffusion at large scales.
The possibility of growing modes with $\lambda > 0$
is obtained if one can have a potential well with
$U$ being sufficiently negative in some 
intermediate range of $r$.
The growth rate $\lambda$
is of the order of the fastest eddy turn over
rate for a sufficiently supercritical $\Rm$ on this scale.
This $\lambda$ then also gives an estimate of
the maximum depth, $U_0$, of the potential, or $U_0 \sim -\lambda$.
The bound state behavior also implies that the magnetic 
correlations die away rapidly for scales larger than the 
correlation scale of the stirring.
\cite{Kaz68} also showed that, for a single scale flow
(or below the viscous cut-off scale in a large $\Pm$ turbulent flow),
the magnetic power spectrum scales as $\EM(k) \propto k^{3/2}$,
until the resistive cutoff scale, $k_\eta \sim \kf \Rm^{1/2}$, where $\kf$
is again the wavenumber of the energy-carrying eddies. 
It turns out that the Kazantsev spectrum is preserved, 
even for a finite correlation time of the velocity field, 
to the lowest order departures from $\delta$-correlated flow \citep{BS14}. 
This \citet{Kaz68} result is generalized in Appendix~\ref{app} to include
the effect of kinetic helicity of the flow.
We will need the resulting asymptotic scaling of $\EM(k)$
in our arguments below.

In the presence of helical velocity correlations $F(r)$, a remarkable
change occurs.
The quantity $\alpha(r\to\infty) = -2F(0) \equiv \alpha_0$ is what is traditionally 
called the $\alpha$ effect. Its presence allows correlations
to grow on scales larger than that of the turbulent velocity field;
i.e. the large-scale magnetic field \citep{Sub99}. 
This can easily be seen from
\Eqs{mlequn}{mhequn}, where even for $r \to \infty$,
we have new generating terms due to the $\alpha$ effect
in the form
$\dot M_{\rm L} = .... + 4\alpha_0 C$ and $\dot H = ... + \alpha_0 M_L$.
These couple $M_{\rm L}$ and $C$
and lead to a growth of large-scale correlations.
Indeed for any quasi-stationary states ($\lambda \sim 0$),
one finds that the problem of determining the magnetic
field correlations
once again becomes the problem of determining
the zero-energy eigenstate
in a modified potential, $U - \alpha^2/\eta_{\rm T}$.
This potential does not go to zero as $r\to \infty$, but instead
tends to a negative definite constant
$-\alpha_0^2/\etaTz$.
So there are strictly no bound states
with zero energy/growth rate, for which the correlations
vanish sufficiently rapidly at infinity; instead the situation is akin to
tunnelling states in quantum mechanics \citep{Sub99,BS00,BCR05,MB07,MB10}.

\subsection{Unified growth of large-and small-scale fields}

In fact, even when $\lambda \ne 0$, like for the 
fastest growing modes with growth rates comparable to
eddy turn over rates, \Eqs{mlequn}{mhequn} can be solved exactly
in the limit $r\to \infty$. 
The solution is most transparent for the correlator 
$w(r,t) = \bra{\BB(\xx,t)\cdot\BB(\yy,t)} 
= M_L + 2M_N$. 
One finds from fairly
straightforward algebra that, for a mode growing with growth 
rate $\lambda$ and scale $r\gg l$
(much larger than the turbulent forcing scales),
\EQ
w(r,t) = e^{\lambda t} \exp(-\ksmall r) 
\frac{A \cos\kmean r + B \sin\kmean r}{r}\;\;\mbox{($r\gg l$)},\;
\label{finsoln}
\EN
where
\EQ
\ksmall = \frac{(2\etaTz\lambda - \alpha_0^2)^{1/2}}{2 \etaTz}, \quad
\kmean = \frac{\alpha_0}{2\etaTz}.
\label{scales}
\EN
Note that this solution applies for real $\ksmall$, or 
$\lambda > \alpha_0^2/2 \etaTz$, that is for growth rates
larger than those of the traditional $\alpha^2$ dynamo
whose maximum growth rate for $\BB^2$ is also $\alpha_0^2/2\etaTz$;
see also \citet{MB07,MB10}.
This would generically apply if strong small-scale dynamo
action is present, that is, when $\Rm$ is large enough and,
in addition, the eddy turn over rate is bigger than the $\alpha^2$ dynamo
growth rate. However, even in this case we see that the presence of the
large-scale field due to the $\alpha$ effect is evident in the
correlator $w(r)$, as reflected in the presence of the oscillating
cosine and sine terms in \Eq{finsoln}. In fact, $\kmean = \alpha_0/2\etaTz$
is exactly the wavenumber for which the growth rate of the
mean field $\alpha^2$ dynamo is maximum; see \Sec{MagneticSpectra}.
This suggests that the
fluctuation dynamo, which is amplifying the field at a rate $\lambda$,
is seeding the simultaneous growth of the large-scale field
with a wavenumber $\kmean$.
In other words, the field in this case
is growing as one eigenfunction such that the large-scale field is
enslaved to the growth of the small-scale field growth.
Such a picture is qualitatively consistent with what is found from
our DNS for large $\Rm$.
We will refer to this as Type~I; see \Tab{Tcomp}.

\begin{table}\caption{
Comparing cases~I and II.
}\vspace{12pt}\centerline{\begin{tabular}{ll}
\hline
\hline
Type~I & SS dynamo dominant \\
       & LS dynamo seeded by it \\
\hline
Type~II & LS dynamo dominant \\
        & SS dynamo enslaved to it \\
\hline
\label{Tcomp}\end{tabular}}\end{table}

The presence of a nonzero $\alpha_0$ can also lead to growth of the
field, even when $\Rm$ is not large enough to excite the small-scale
dynamo. In this situation the $\alpha^2$ dynamo can be
excited, with a continuous spectrum
of eigenmodes with $\lambda \le \alpha_0^2/2\etaTz$.
The eigenfunction for large $r\gg l$
(i.e.\ for scales much larger than the turbulent forcing scales)
then changes to
\EQ
w(r,t) = e^{\lambda t} 
\frac{A \cos\bar\kmean r + B \sin\bar\kmean r}{r}\quad\mbox{($r\gg l$)},
\label{finsoln2}
\EN
where
\EQ
\bar\kmean = \frac{\alpha_0 - (\alpha_0^2 - 2\lambda \etaTz)^{1/2}}
{2\etaTz}.
\label{scales2}
\EN
Again the presence of the large-scale field is evident
due to the cosine and sine terms in the correlator $w(r)$.
The fastest growing mode in this case
has $\lambda=\alpha_0^2/2\etaTz$ and $\bar{k}_{\rm m} \equiv k_{\rm m}$.
Moreover, for these solutions the small-scale fields
on scales $r <l$ are enslaved to the large-scale dynamo
and arise by the velocity field tangling up the large-scale field.
Such a solution is what one obtains in our DNS at small $\Rm < \Rmc$.
We refer to this case where the large-scale dynamo is dominant as Type~II;
see \Tab{Tcomp}.

\subsection{$\Rm$ dependence of large-scale field strength}

The other question is why the large-scale field strength,
as measured by the ratio $\meanB/\Brms$, decreases with $\Rm$? 
It is also somewhat surprising that the large-scale field,
decreases with $\Rm$ even for
moderate $\Rm < 100$, especially for $\Pm=0.1$
when one does not naively expect the small-scale dynamo
to operate \citep{Isk07} (see also \Fig{pRm_crit}).
There are potentially two effects.
First there could be a decrease of 
the strength of the eigenfunction, at the scale $r \sim \kmean^{-1}$, 
compared to the forcing scale $l \sim \kf^{-1}$.
This is obtained for Type~I, where the small-scale dynamo
operates, with the large-scale field enslaved to it. Here 
due to the $\exp(-\ksmall r)$ term in \Eq{finsoln},
the strength of the eigenfunction, at the scale $r \sim \kmean^{-1}$, 
would have decreased exponentially by a factor $\sim \exp(-\ksmall/\kmean)$
from its value at smaller scales.
As the ratio $\ksmall/\kmean$
increases with increasing growth rate $\lambda$,
which itself increases with $\Rm$ in our simulations,
one can obtain a smaller mean field 
compared to the field at the forcing scale, with increasing $\Rm$.

This effect is however not present when the large-scale dynamo 
is dominant, as the  $\exp(-\ksmall r)$ term is absent in this case 
(see \Eq{finsoln2}).
There is however a second effect which is likely to be the more
dominant one at large $\Rm$ during the kinematic stage
in both Types~I and II. This is obtained, as we show below, due to
the fact that the magnetic power spectrum generically
increases further from the forcing wavenumber $\kf$ 
to peak at the resistive one $k_\eta$, which itself
increases with $\Rm$.
We discuss this below.

Note that for a purely non-helical small-scale dynamo, the magnetic spectrum
in the kinematic stage is expected to increase as $\EM(k) \propto k^s$
from the forcing scale $\kf$ to the $\Rm$-dependent
resistive scale, say $k_\eta$.
In case of a single scale flow, one has the Kazantsev spectrum with, 
the spectral index $s=3/2$. What happens when helicity is included,
and large-scale field generation becomes possible?

The influence of helicity on the large $k$ behaviour
of the magnetic spectrum, for large $\Rm$, 
is analyzed in some detail in Appendix~\ref{app}.
In particular, we consider
the coupled system given by 
\Eqs{mlequn}{mhequn} on scales that are much
larger than the resistive scale, but much smaller than the
outer forcing scale $l$ of the random motions of the turbulence.
In this range one can approximate
$\eta_{\rm t}(r)$ and $\alpha(r)$ as power laws.
We show quite generally that even in case of helical flows, 
where large-scale dynamo
action is in principle possible, the magnetic spectrum at the kinematic
stage is peaked at resistive scales. 

Surprisingly, for both a single-scale flow, and for 
Kolmogorov scaling of the velocity spectra,
with maximal kinetic helicity at the forcing scale,
we find that helicity is unimportant for the
behaviour of the magnetic spectrum at large $k$.
For a helical single scale flow, the
magnetic spectrum still scales as the Kazantsev spectrum, 
$\EM(k) \propto k^{3/2}$ at large $k$.
For Kolmogorov scaling of the velocity spectra,
with maximally helical forcing,
we show in \App{app} that the magnetic spectrum is still peaked
at resistive scales; and at large $k$ it
is of the form $\EM(k) \propto k^s$ with $s \approx 7/6$.
Thus, for the kinematic dynamo, even though large-scale
fields are being generated due to the presence of
helicity, the magnetic power spectrum is still peaked
at resistive scales.
Our DNS also suggest such the conclusion that
$\EM(k) \propto k^s$, with $s>0$, as can be
seen from the spectra shown in 
\Figs{ppspecmtfit_comp}{ppspecmtfit_comp_256Pm1a}.
Note also that these conclusions are quite independent
of whether the dynamo is predominantly
a large-scale or small-scale dynamo,
and only depends on there being scale separation between the forcing
and resistive scales, as one would obtain for sufficiently large $\Rm$.
We can now ask what this implies for the behaviour of $\meanB/\Brms$,
with $\Rm$?

Now suppose the magnetic power spectrum increases with $k$ as
$\EM(k) \propto k^s$ for $\kf < k < k_\eta$ and $k_\eta \propto \Rm^\beta$.
Integrating the spectrum over $k$ from $\kf$ to $k_\eta$, we find for the
ratio $(\Brms/\Bf)^2 \propto (k_{\eta}/\kf)^{s+1} \propto \Rm^{\beta(s+1)}$,
where we have defined the small-scale field at the forcing scale as
$\Bf = (\kf M(\kf))^{1/2}$.
Thus, $(\Brms/\Bf) \propto \Rm^{\beta(s+1)/2}$.
For a single-scale flow we have
$s=3/2$ and $\beta=1/2$, and then $\Brms/\Bf \propto \Rm^{5/8}$.
On the other hand, for Kolmogorov scaling of the velocity spectra
with $s=7/6$ and, say, $\beta=3/4$, we have $(\Brms/\Bf) \propto \Rm^{0.81}$
scaling. At the same time, we have seen that 
$\meanB/B_s \sim \exp(-\ksmall/k_m)$
for Type I with $B_s \sim \Bf$.
For Type II, where the large-scale dynamo dominates,
one would expect the rms value of $\meanBB$ to be comparable to
$\Bf$, as would be the case when there is a $k^{-1}$ spectrum
\citep{RS82} between $k_m$ and $k_f$.
Combining these arguments, we do expect $\meanB/\Brms$ to decrease
significantly with $\Rm$, although the exact scaling as $\Rm^{-1/2}$,
or the further scaling as $\Rm^{-3/4}$, are not yet fully understood.

\section{Conclusions}
\label{Conclusions}

We have shown here that large-scale dynamo action is obtained
in large $\Rm$ helical turbulence in the kinematic stage, even when a 
strong small-scale dynamo is also possible.
Both large and small scales grow at the same rate, such
that the energy spectrum is shape invariant in the kinematic stage.
By splitting the magnetic energy spectrum into positively and negatively
polarized parts, $\EMz^\pm$, clear signatures of large-scale fields can be
seen at small $k$ as an excess power in $\EMz^-(k)$ ($\EMz^+(k)$) if the
kinetic helicity at the forcing scale is positive (negative).
Evidence for the large-scale mean field $\meanB$ is also clearly seen in 
suitably defined planar averages. 
This evidence for a mean field in helically driven
turbulence is as expected for the standard $\alpha^2$ mean-field dynamo, and
thus allows us to prove the existence of such a mean-field dynamo effect.

The DNS also show that both the amplitude of the large-scale field 
and the dynamo growth rate increase with increasing fractional helicity. 
This is as expected and helps to determine the onset
of large-scale dynamo action and to distinguish it from that of the
small-scale dynamo.
As a by-product of our work, we find that
for $\kf/k_1=4$, the $\Rmc$ for exciting the 
small-scale dynamo at small $\Pm$ is different from earlier
results which were based on smaller scale separation, $\kf/k_1=1$--$2$.
For example, the threshold magnetic Reynolds number for $\Pm=0.1$ is
decreased to a modest value of $\Rmc \approx 160$.

The mean field found from the DNS using planar averages, however decreases with
$\Rm$ as $\Rm^{-1/2}$ (or possibly faster) in the kinematic stage.
Such a decline is obtained both when the small-scale dynamo is dominant
(Type~I) and also when the large-scale dynamo is dominant,
but the small-scale dynamo enslaved to it (Type~II).
By analyzing the Kazantsev model including helicity,
this feature is shown to arise due to the fact that the
magnetic spectrum $\EMz(k)$ for large $\Rm$, is peaked at the
resistive scale, even when helicity is present. 
Such a rise in $\EMz(k)$ with $k$ is also seen in the DNS
that we have performed.
 
This raises the question, does kinematic dynamo theory have any relevance?
The answer is yes, because it allows us to {\em identify} mechanisms
that may have a connection with the nonlinear regime where the
large-scale dynamo becomes dominant and the
small-scale power is lost (mode cleaning).
Firstly, nonlinear simulations of the small-scale dynamo at large $\Rm$, 
which have a large enough inertial range
show that the nonlinear evolution can lead to a significant
increase in the magnetic integral scale \citep{HBD04,CR09,Eyink13,BS13}.
Thus, the effect of the Lorentz force is to bring the power from the
resistive scale to scales just smaller than the forcing scale.
Also, simulations of the $\alpha^2$ dynamo in periodic domains,
show that the magnetic field becomes ordered
on the largest available scales, independently of $\Rm$, provided
small-scale magnetic helicity can be dissipated \citep{B01,B09,CB13}.
Therefore, the combined action of the Lorentz force to transfer power from
resistive scales to larger scales, and small-scale helicity loss from
the system, could result in an efficient generation of
the large-scale field, even in the presence of the fluctuation dynamo.

For the transfer of power from resistive scales to larger scales
to happen, the spectrum must change shape during saturation
such that large spatial scales (small $k$) 
can still be amplified while small scales (large $k$) saturate.
Recall that all scales grow at the same rate during the kinematic stage.
In terms of the potential picture of the Kazantsev model with helicity 
(\Secs{intro}{unified}), the
potential well at the small scale $l$ needs to become shallower 
due to nonlinear effects to allow for only the marginally bound state 
to exist, while still having sufficient depth at the large scale $L$,
to allow the `tunnelling free particle' states to grow. Such local saturation
in a related real-space double well potential problem has been found
in the context of a spirally forced nonaxisymmetric galactic dynamo 
\citep{CSS13a,CSS13b}.
There the potential wells are near the galactic centre and
the corotation radius of the spiral, so the eigenfunction grows 
fastest in the central regions, with its tail seeding the growth of the 
nonaxisymmetric magnetic spiral field around corotation.
Saturation of the dynamo near the galactic centre still allows for the
field to grow around corotation and become significant.
Whether such a situation can also be obtained for a double well potential 
in `scale' or wavenumber space remains to be determined.
It would be of interest to verify this 
in a nonlinear version of the Kazantsev model, where helicity loss
can also be built in, and perhaps even more importantly, in high resolution
DNS which can resolve both the small-scale dynamo and have
enough scale separation to simultaneously capture the large scales.

\section*{Acknowledgements}
We thank the referee for useful comments which have led to
an improvement of the paper.
KS thanks Nordita for hospitality during his
visit there, which led to the present work. 
This work was supported in part by
the European Research Council under the AstroDyn Research Project No.\ 227952,
and the Swedish Research Council grants No.\ 621-2011-5076 and 2012-5797,
as well as the Research Council of Norway under the FRINATEK grant 231444.
We acknowledge the allocation of computing resources provided by the
Swedish National Allocations Committee at the Center for
Parallel Computers at the Royal Institute of Technology in
Stockholm and the National Supercomputer Centers in Link\"oping,
the High Performance Computing Center North in Ume\aa,
and the Nordic High Performance Computing Center in Reykjavik.


\appendix

\section{The influence of helicity on small scales}
\label{app}

The purpose of this appendix is to
analyze the behaviour of the coupled system given by 
\Eqs{mlequn}{mhequn}, on scales that are much
larger than the resistive scale, but much smaller than the
outer forcing scale $l$ of the random motions or the turbulence.
In this range one can approximate
$\eta_{\rm t}(r)$ and $\alpha(r)$ as power laws.
We take quite generally
\EQ
\eta_{\rm t}(r) = \etaTz \left(\frac{r}{l}\right)^q \quad {\rm and} \quad 
\alpha(r) = \alpha_0 \left(\frac{r}{l}\right)^p.
\label{etaalpr}
\EN
For a single scale flow we adopt $p=q=2$.
For a Kolmogorov spectrum $E(k) \propto k^{-5/3}$, 
we can use Richardson scaling
for the scale-dependent turbulent diffusion and take
$q = 4/3$ \citep{Vain82}.
Suppose further that the flow is driven by a fully helical forcing.
Then \citet{BS05b} found that the kinetic helicity spectrum 
also scales as $\HK(k) \propto k^{-5/3}$.
Therefore $\alpha(r=1/k) \propto \tau(k) (k\HK(k)) \propto r^{4/3}$,
where $\tau(k) \propto k^{-2/3}$ is a scale-dependent correlation time.
Thus, for a Kolmogorov energy spectrum, assuming also a fully helical
velocity field, one could adopt $q=4/3$, $p=4/3$.
We will discuss both cases below.

Let us define a dimensionless coordinate $z=r/l$, adopt the power law forms
given in \Eq{etaalpr}, and look at eigenmode solutions to 
\Eqs{mlequn}{mhequn} of the form
$M_L = \exp(\lambda t) \tilde{M}_L(r)$ and $C = \exp(\lambda t) \tilde{C}(r)$.
We get
\EQA
\bar\lambda \tilde{M}_L(z)&=&
\left(\frac{\eta}{\etaTz} + z^q\right)\tilde{M}_L^{''}
+\left(\frac{4\eta}{\etaTz} + (4+q) z^q\right) \frac{\tilde{M}_L^{'}}{z}
\nonumber \\
&+& q(3+q) z^{q-2}\tilde{M}_L + 4\bar\alpha_0 z^p \tilde{C}(z),
\label{tildeML}
\ENA
\EQA
\bar\lambda \tilde{C}(z)&=&
\left(\frac{\eta}{\etaTz} + z^q\right)\tilde{C}^{''}
+\left(\frac{4\eta}{\etaTz} + (4+2q) z^q\right) \frac{\tilde{C}^{'}}{z} 
\nonumber \\
&+& q(3+q) z^{q-2}\tilde{C} -
\bar\alpha_0 z^p \tilde{M}_L^{''}
\nonumber \\
&-&2\bar\alpha_0 (p+2) z^{p-1} \tilde{M}_L^{'}
- \bar\alpha_0 p(p+3) z^{p-2} \tilde{M}_L.
\label{tildeC}
\ENA
Here we have defined a dimensionless growth rate
$\bar\lambda = l^2\lambda/(2\etaTz)$.
In the limit $z^q \gg \eta/\etaTz$, 
or $z \gg z_\eta = (\eta/\etaTz)^{1/q}$, one can neglect the
resistive terms in \Eqs{tildeML}{tildeC}. 
(Here $z_\eta$ is the dimensionless resistive scale.)
Note that without the mutual coupling due to 
the $\alpha$ effect, these equations would be scale
free in the sense that a transformation of
$z\to c z$ leaves \Eqs{tildeML}{tildeC} invariant.
The question arises if there still exist scale-free
solutions in the presence of an $\alpha$ effect.
As power laws are scale free, we examine if
\Eqs{tildeML}{tildeC} can have power law solutions
of the form say $\tilde{M}_L = M_0 z^{-\mu}$, $\tilde{C} = C_0 z^{-\nu}$.
Substituting this form for $\tilde{M}_L$ and $\tilde{C}$ gives
\EQA
\bar\lambda M_0 &=& \left[\mu(\mu+1)  - \mu(4+q)  +q(3+q)\right]M_0 z^{q-2} 
\nonumber \\
&+& 4\bar\alpha_0 C_0z^{p +\mu-\nu},
\label{M0eqn}
\ENA
\EQA
\bar\lambda C_0 &=& \left[\nu(\nu+1) - \nu(4+2q) +q(3+q)\right]C_0 z^{q-2}
\nonumber \\
&-& \bar\alpha_0 M_0
z^{p-2+\nu-\mu}
\left[\mu(\mu+1) - \mu(4+p) +p(3+p)\right].
\nonumber \\
\label{Ceqn}
\ENA
Thus, a scale-free solution can be obtained if the $z$ dependence
drops out in \Eqs{M0eqn}{Ceqn}. To see if this can be obtained, 
consider now the two cases which we mentioned above.
In the case of a single-scale flow with $p=q=2$, we have
$q-2=0$, and the first terms on the right-hand side of 
\Eqs{M0eqn}{Ceqn} become $z$ independent. On the other hand, 
the exponent of $z$ in the
last term of \Eq{M0eqn} becomes $\mu-\nu+2$, while that 
in \Eq{Ceqn} becomes $\nu-\mu$.

One can get a nearly scale invariant solution if
$\mu=\nu$, which implies that \Eq{Ceqn} becomes $z$-independent, 
while $\mu-\nu+2=2$ in \Eq{M0eqn}.
Then the exponent of $z$ in the
last term in \Eq{M0eqn} becomes $2$ and the $z$-dependent term in
\Eq{M0eqn} is $\propto z^2 \ll 1$, and thus can be neglected.
In this case, the helical part of the correlation
completely decouples from the non-helical part of the correlation.
\EEq{M0eqn} then reduces to that obtained for the
standard non-helical small-scale dynamo \citep{Kaz68,BS14}, and one
recovers the Kazantsev spectrum, $\EM(k) \propto k^{3/2}$.
Thus, even in the presence of helicity in the velocity field, 
if the fastest growing mode is being driven effectively by 
a single scale flow, then helicity is unimportant for the
behaviour of the magnetic spectrum at large $k$!

The nature of the small $z$ (or large $k$) solution can be explicitly seen by
looking at the solution to the resulting quadratic
equation for $\mu$ given by \Eq{M0eqn}; cf.\ \cite{BS14}.
We get for $\mu$
\EQ
\mu^2- 5\mu + (10-\bar\lambda) = 0,
\quad {\rm so} \ \mu = \frac{5}{2} \pm i \mu_I,
\label{smallrSS}
\EN
where $\mu_I = [4(10-\bar\lambda) -25]^{1/2}/2$ can be shown to be small
(once $\bar\lambda$ is determined),
and importantly, the real part of $\mu$ is $\mu_R=5/2$.
From \Eq{smallrSS}, in the range 
$z_\eta \ll z \ll1$,
$M_L$ is then given by
\EQ
M_L(z,t) = e^{\gamma\tilde{t}} \tilde{M}_0
z^{-\mu_R} \cos\left(\mu_I \ln z + \phi\right),
\label{solnml}
\EN
where $\tilde{M}_0$ and $\phi$ are constants.
Thus $M_L$ varies dominantly as $z^{-5/2}$,
modulated by the weakly varying cosine factor (both because
the phase of the cosine depends on the weakly varying
$\ln z$ and because $\mu_I$ is small).
The magnetic
power spectrum is related to $M_L$ by
\EQ
\EM(k,t) = \int dr (kr)^3 M_L(r,t) j_1(kr). 
\label{MkMl}
\EN
The spherical Bessel function $j_1(kr)$ is peaked around
$ k\sim 1/r$, and a power law 
behaviour of 
$M_L \propto z^{-\lambda_R}$,
for $z_\eta \ll z \ll l$,
translates into a power law
for the spectrum $\EM(k) \propto k^{\lambda_R -1}$ 
at large $k$ (but smaller than the resistive scale, i.e.\
with $k_\eta = l/z_\eta \gg k \gg 1/l$). 
As $\lambda_R =5/2$ for a single-scale flow, 
this implies that the magnetic spectrum is of the Kazantsev form 
with $\EM(k) \propto k^{3/2}$ in $k$ space, as advertised above.

Note that, although the $\alpha$-effect does not affect the magnetic
energy spectrum at large $k$ for a single scale flow, 
it is indeed important in driving the
current helicity evolution. 
From \Eq{Ceqn}, we get $C_0 = \bar\alpha_0 M_0(1 - \bar\lambda/2\mu)$,
which can be used to write $C(r,t)$ explicitly.
(We note in passing that the other potentially scale-invariant 
case would have $\mu-\nu+2=0$. This however implies
$\nu = \mu +2$, and turns out to violate the realizability
condition, which in real space requires $\nu \le \mu +1$,
for power law correlations/spectra.)

Now consider the other case of Kolmogorov scaling with $q=4/3$, $p=4/3$.
In this case, the first terms on the right-hand side of
\Eqs{M0eqn}{Ceqn} are proportional to $z^{-2/3}$.
On the other hand, the exponent of $z$ in the
last term of \Eq{M0eqn} becomes $4/3 + \mu-\nu$, while that 
in \Eq{Ceqn} becomes $-2/3 +\nu-\mu$.
Multiplying both \Eqs{M0eqn}{Ceqn} by $z^{2/3}$, we have
\EQA
\bar\lambda M_0 z^{2/3} &=& \left[\mu(\mu+1)  - \mu(4+q)  +q(3+q)\right]M_0 
\nonumber \\
&+& 4\bar\alpha_0 C_0z^{2 +\mu-\nu}
\label{M0eqnk}
\ENA
\EQA
\bar\lambda C_0 z^{2/3} &=& \left[\nu(\nu+1) - \nu(4+2q) +q(3+q)\right]C_0
\nonumber \\
&-& \bar\alpha_0 M_0
z^{-(\mu-\nu)}
\left[\mu(\mu+1) - \mu(4+p) +p(3+p)\right].
\nonumber \\
\label{Ceqnk}
\ENA
Now, for $z\ll1$, the left-hand side of the above equations
will be small and can be neglected.
One can then again get a nearly scale invariant solution if
$\mu=\nu$, 
which implies that the right hand side of \Eq{Ceqnk} becomes $z$-independent, 
while $\mu-\nu+2=2$ in \Eq{M0eqnk}.
Then the exponent of $z$ in the
last term in \Eq{M0eqnk} again becomes $2$ and the $z$-dependent term in
\Eq{M0eqnk} is $\propto z^2 \ll 1$, and thus can be neglected.
In this case, just as in the case of a single scale flow, 
the helical part of the correlation
completely decouples from the non-helical part of the correlation
at small $z$, in \Eq{M0eqnk}. 
The condition that the resulting homogeneous equation 
for $M_0$ has nontrivial solution implies
\EQ
\mu^2 -\frac{13}{3}\mu +\frac{52}{9} = 0
\label{kolcase}
\EN
The resulting quadratic equation has complex conjugate roots,
$\mu = \mu_R \pm i \mu_I$, where now $\mu_R = 13/6$ and 
$\mu_I =\sqrt{39}/6$, correspond to the 
solution for $M_L$ given in \Eq{solnml}.
Although $\mu_I$ is now larger and the cosine factor
in \Eq{solnml} varies by a larger factor, the
power law envelope $M_L \propto z^{-\mu_R} \propto z^{-13/6}$
now corresponds to an approximate spectral dependence $\EM(k) \propto k^{7/6}$
at large $k$.
 
In summary, even in the case of helical flows, where large-scale dynamo
action is in principle possible, the magnetic spectrum at the kinematic
stage is peaked
at resistive scales, with $\EM(k) \propto k^s$ at large $k$, 
where $s$ ranges from $3/2$ (for single scale flow) to about 
$7/6$ for Kolmogorov scaling of the velocity spectra.

\end{document}